\documentclass[aps,prd,twocolumn,superscriptaddress,nofootinbib,eqsecnumm,showpacs]{revtex4}
\usepackage{graphicx,longtable,mathrsfs,color,array}
\usepackage[hidelinks]{hyperref}
\usepackage[usenames,dvipsnames]{xcolor} 
\usepackage{amssymb,amsmath,mathtools,mathrsfs,slashed} 
\usepackage{epsfig,subfigure,placeins,float} 
\usepackage{booktabs,longtable,ctable,multirow} 
\usepackage{exscale,relsize} 
\usepackage[normalem]{ulem} 
\usepackage{enumerate}
\usepackage{times, mathptmx} 
\usepackage[utf8]{inputenc}
\usepackage{color}
\usepackage{hyperref}
\usepackage{graphicx}
\usepackage{color}
\usepackage{placeins}
\usepackage{graphicx,graphics}
\usepackage{tabularx}
\allowdisplaybreaks[1]

\newcommand{\be}{\begin{equation}}
\newcommand{\ee}{\end{equation}}
\newcommand{\beq}{\begin{equation}}
\newcommand{\eeq}{\end{equation}}
\newcommand{\bea}{\begin{eqnarray}}
\newcommand{\eea}{\end{eqnarray}}

\definecolor{dgreen}{rgb}{0,0.7,0.0}

\begin{document}

\title{Gravitational wave propagation beyond geometric optics}

\author{Giulia Cusin}
\email{giulia.cusin@physics.ox.ac.uk}
\affiliation{Astrophysics, University of Oxford, DWB, Keble Road, Oxford OX1 3RH, UK}
\author{Macarena Lagos}
\email{mlagos@kicp.uchicago.edu}
\affiliation{Kavli Institute for Cosmological Physics, The University of Chicago, Chicago, IL 60637, USA}
\date{Received \today; published -- 00, 0000}

\begin{abstract}
It is standard practice to study the lensing of gravitational waves (GW) using the geometric optics regime. However, in many astrophysical configurations this regime breaks down as the wavelength becomes comparable to the Schwarzschild radius of the lens. We revisit the lensing of GW including corrections beyond geometric optics. We propose a perturbative method for calculating these corrections simply solving first order decoupled differential equations. We study the behaviour of a single ray and find that the polarization plane defined in geometric optics is smeared due to diffraction effects, which leads to the rise of apparent vector and scalar polarization modes. We analyze how these modes depend on the observer choice, and we study the impact of diffraction on the  pseudo-stress energy momentum tensor of the gravitational field.
 \end{abstract}
\keywords{Black holes, Lensing, Gravitational Waves}

\maketitle

\section{Introduction}


The direct detection of gravitational waves (GW) by the LIGO/Virgo collaboration \cite{LIGO, 2015CQGra..32b4001A} has given us a new way of observing the cosmos. Instead of measuring electromagnetic waves at various frequencies, we can detect perturbations of the spacetime itself. Very much like the electromagnetic field, these perturbations obey a wave equation -- hence being dubbed gravitational waves. 

In studying the propagation of gravitational waves from distant objects, the standard approach is to use the geometric optics regime. In practice, this means that waves propagate along geodesics of the background (or perturbed) spacetime. Furthermore, the two transverse tensor polarizations of the wave are unchanged apart from parallel transport along the geodesic path. This means that effects such as lensing or time delay can be calculated for gravitational waves much in the same way as is done for light rays. One can then import the techniques that have been developed for light propagation in astrophysics and cosmology (such as, for example, the cosmic microwave background or galaxy lensing) directly into gravitational wave physics (see e.g.~\cite{Laguna_2010, Camera:2013xfa, BERTACCA201832, Mukherjee:2019wcg} for weak lensing analyses of GW).

\begin{figure}
\centering
\includegraphics[scale=0.36]{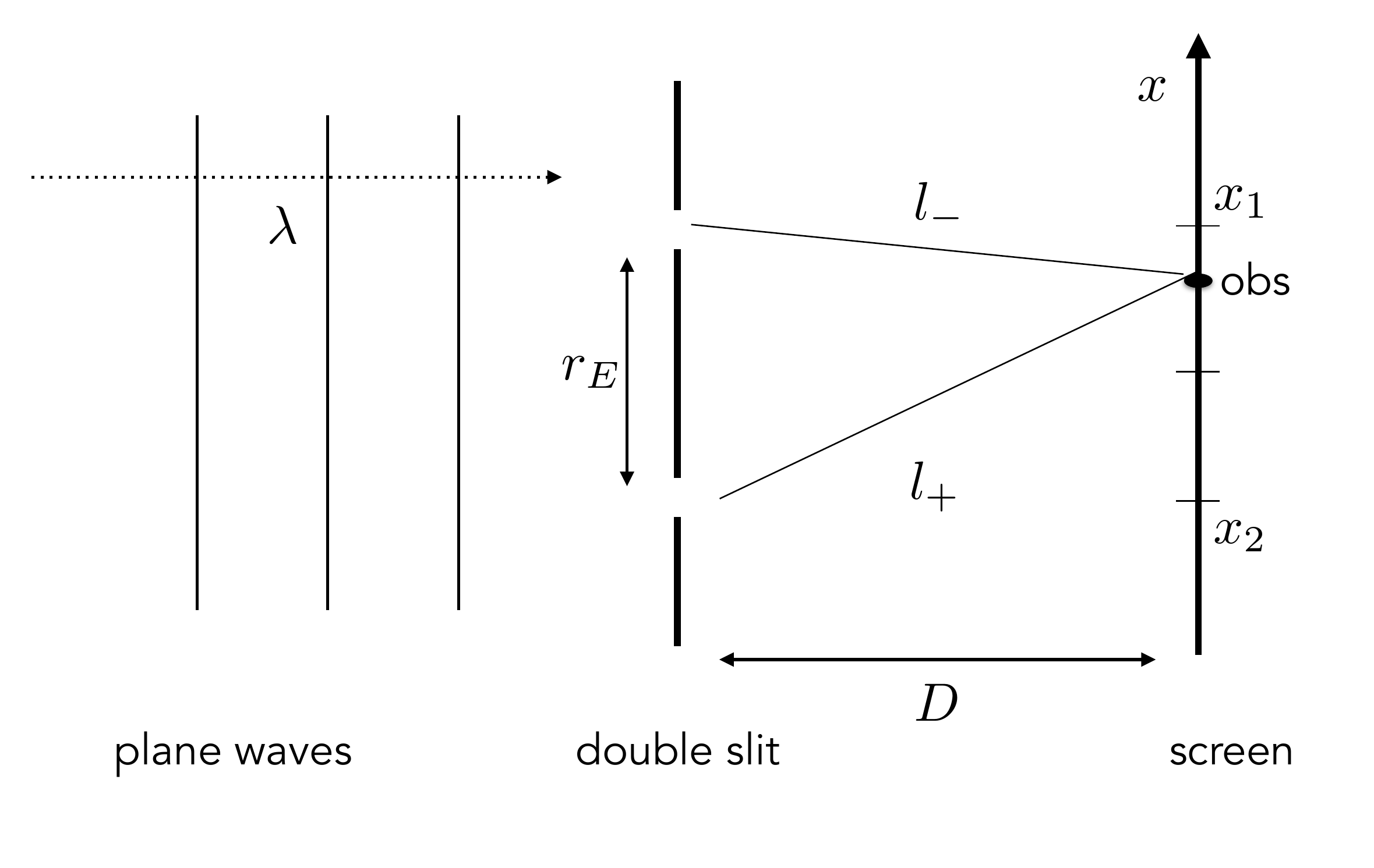}
\caption{\small \label{double slit} Schematic representation of the double slit experiment.}\end{figure}

If one scrutinizes the conditions under which geometric optics can be applied, one finds that it may not be appropriate in the case of realistic scenarios for gravitational waves \cite{1998PhRvL..80.1138N, Takahashi:2003ix, Macquart:2004sh,Takahashi:2016jom}. 
Consider a localized object with a certain mass (or equivalently, Schwarzschild radius) acting as a lens for an incoming monochromatic plane wave with wavelength $\lambda$. 
The geometric optics approximation is valid in the weak field regime when the wavelength is much smaller than the Schwarzschild radius of the lens \cite{1998PhRvL..80.1138N, Takahashi:2003ix, Macquart:2004sh,Takahashi:2016jom}. We can understand this by making an analogy with the double slit experiment, with the geometry represented in Figure \ref{double slit}. When waves with wavelength $\lambda$ pass through the slits, an interference pattern is produced on the screen. We denote  the distance from each slit to the observer as $l_{\pm}$. By geometry, these distances can be expressed as:
\be
l_{\pm}=\sqrt{(x\pm r_E/2)^2+D^2}\,,
\ee
where $r_E$ is the separation between the slits, $D$ is the distance between the slits and the screen, and $x$ is the position of the observer on the screen, measured from the point on the screen with the minimum distance to the middle point between the two slits. Denoting the path length differences as $\Delta l=|l_+-l_-|$ we obtain
\be
\Delta l\simeq \frac{r_E x}{D}\,,
\ee
where we have assumed $D\gg (r_E, x)$. The width of the central peak is obtained setting $\Delta l\sim \lambda$ and it is given by
\be\label{deltax}
\Delta x\sim \frac{D}{r_E} \lambda\,.
\ee
In the ray optics regime, the radiation incoming across the screen effectively behaves as a particle. On the screen, only observers located in correspondence to the two slits in $x_1$ and $x_2$ would receive a signal. In particular, an observer located at $x=0$ would not receive any signal, i.e.~the width of the first peak can be neglected in this regime. In more formal terms, we can state that the regime of validity of ray optics is when $\Delta x \ll r_E$, i.e.~using (\ref{deltax}), for 
\be\label{condition}
\lambda\ll r_E^2/D\,.
\ee

We can now think of a similar setting, where instead of the two slits we have a point-like lens with Einstein radius \cite{1992grle.book.....S}
\be
r_E=\sqrt{2 r_S \frac{D_{OL}D_{LS}}{D_{OS}}}\,,
\ee
where $D_{LS}$, $D_{OS}$, and $D_{OL}$ are the angular diameter distances of  lens-source, observer-source, and observer-lens, respectively. 
Then it turns out that the validity of ray optics to study the propagation of the wave after the lens is given by (\ref{condition}). Assuming for simplicity $D_{OL}\sim D_{LS}\sim D_{OS}$,  this gives back the condition that the wavelength is smaller than the Schwarzschild radius of the lens, $\lambda\ll r_S$. This is a necessary condition for geometric optics to be valid.\footnote{Note that the condition is actually given by $\lambda \ll 2r_S D_{LS}/D_{OS}$. The geometric factor $D_{LS}/D_{OS}$ is of order one when the lens is close to the observer and it decreases as the lens gets closer to the source. However, it gets significantly small only when the lens is very close to the source. As an example, for a source at 40 Mpc from us and a lens at 38 Mpc (i.e.~much closer to the source than to us) in a $\Lambda$CDM universe,  the condition of validity of geometric optics reads $\lambda \ll 0.1  r_S$. Nevertheless, we will neglect this geometric factor when making order of magnitude estimates throughout the paper.}

Geometric optics is certainly appropriate for waves from gamma to radio frequencies, but that may not be the case for gravitational waves. Indeed, Pulsar Timing Arrays\footnote{See e.g. {\tt http://www.ipta4gw.org}} and the Laser Interferometer Space Antenna (LISA)\footnote{{\tt www.lisamission.org}} may detect GW with wavelength of astrophysical lengths, of the order of fraction of parsecs. This means that GW will have wavelengths comparable or even larger than the Schwarzschild radius of astrophysical objects, and standard lensing processes will likely include wave effects that must be appropriately taken into account.  Furthermore, lens objects that are in optically thick regions (such as inside galaxies) can still be detected by lensing of GW since GW propagate through surrounding matter without much absorption.  As a consequence, GW can probe much smaller  lenses, at the sub galactic scale.

We emphasize that the analysis of wave effects is essential in order to appropriately interpret the GW signals received, reconstruct the gravitational potential of structures along the line of sight, and hence extract unbiased intrinsic properties of GW sources.
Over the years,  a lot of work has been dedicated to the computation of wave effects in GW propagation \cite{1968PhRv..166.1263I, Isaacson:1968zza,1974PhRvD...9.2207P,1976PhRvD..13..775P, 1992grle.book.....S, Cusin:2018avf}, including diffraction, refraction and scattering. 
Regarding lensing of GW, wave effects from binary compact objects have been considered in \cite{1996PhRvL..77.2875W, 1998PhRvL..80.1138N, Takahashi:2003ix, Takahashi:2016jom, Cremonese:2018cyg}. However, in all these works the spin nature of GW is neglected, and the wave is treated as a scalar (spin-0) wave. In other words, it is assumed that the polarization tensor of the wave stays constant during propagation. Indeed, perturbative approaches similar to the one we will consider in this paper has been considered in the past to estimate magnifications and time delays beyond geometric optics \cite{Takahashi:2004mc}. However, in this work, we do not introduce this assumption and we keep track of the spin-2 nature of GW, which allow us to analyze wave effects on both amplitude and polarization. We find that, beyond geometric optics, an incident ray is diffracted and, as a result, the original polarization plane is smeared and effective vector and scalar polarizations arise. 
The work proposed here provides a first step towards disentangling effects coming from propagation in a universe with structures and effects coming from intrinsic properties of the emitting sources. 

Whenever there is a moderate separation of scales (as in most cases of physical interest), one expects that geometric optics  remains a  valid approximation, and that more accurate results can be obtained by including higher-order corrections, which will provide insights into wave-optical phenomena that are not taken into account in the eikonal limit\footnote{The regime we want to describe is similar to the Fresnel regime of optical diffraction (in the absence of curvature), which arises when a wave propagating in an inhomogeneous medium manifests a modest wavefront spreading (and geometric optics is beginning to work, at least roughly).}. This is the idea that we follow in this work: we develop a perturbative approach to study small effects beyond geometric optics. This approach has been used for the case of electromagnetic waves in  \cite{1976JMP....17..576A} and more recently in \cite{Dolan:2018ydp}, where it was shown that corrections beyond geometric optics lead to the so-called spin Hall effect \cite{Oancea:2019pgm}. A similar perturbative approach has been considered in the context of electromagnetic and gravitational wave propagation in \cite{Harte:2018wni}. In this paper,  we  introduce a new framework to recursively solve the beyond geometric optics equations in terms of a system of decoupled first order differential equations. This allows us to investigate general properties of the polarization tensor of the wave as well as its pseudo energy momentum tensor once beyond geometric optics corrections are accounted for.  

This paper is organized as follows. In Section \ref{formalism} we derive the equations describing the propagation of gravitational waves once geometric optics corrections are added. We then perform a perturbative expansion that allows us to identify the geometric optics limit and leading corrections beyond it.
In Section \ref{computation} we propose a framework to explicitly solve these equations in terms of a system of decoupled  first order differential equations. 
In Section \ref{polarization} we introduce the Newman-Penrose scalars, which provide gauge-invariant quantities describing the possible polarizations of GW. 
In Section \ref{propofenergy} we compute the pseudo energy-momentum tensor of the gravitational field and we identify a propagation vector effectively giving the direction of propagation of the wave's average energy. Finally, in Section \ref{discussion} we discuss the regime in which our formalism may be useful and lay out future steps in developing this machinery to accurately predict and asses the detectability of wave effects of GW.
		
Throughout this paper we will use natural units with $c=8\pi G=1$ and signature mostly plus for the metric. We will also denote symmetrization and antisymmetrization of an operator $A_{\mu\nu}$ as $A_{(\mu\nu)}\equiv\frac{1}{2}(A_{\mu\nu}+A_{\nu\mu})$ and $A_{[\mu\nu]}\equiv\frac{1}{2}(A_{\mu\nu}-A_{\nu\mu})$, respectively. 

\section{Formalism}
\label{formalism}
We start by considering the Einstein equations of motion in the presence of matter:
\begin{equation}
G_{\mu\nu}=R_{\mu\nu}-\frac{1}{2}g_{\mu\nu}R=T_{\mu\nu}\,,
\end{equation}
where $G_{\mu\nu}$ is the Einstein tensor and $T_{\mu\nu}$ the stress-energy momentum tensor of matter. In order to study gravitational waves, we separate the dynamical spacetime into a smooth (or slowly varying in space) background field    $\bar{g}_{\mu\nu}$ and a quickly varying perturbation $h_{\mu\nu}$ as:
\begin{equation}
g_{\mu\nu}=\bar{g}_{\mu\nu}+h_{\mu\nu},\quad  |\bar{g}_{\mu\nu}|\gg|h_{\mu\nu}|\,,
\end{equation}
where $h_{\mu\nu}$ describes small ripples on the background, hence identified as gravitational waves. The background metric $\hat{g}_{\mu\nu}$ satisfies the Einstein equations, whereas the linear perturbation satisfies the following dynamical equation:
\begin{eqnarray}
	\psi_{\mu\nu; \alpha}{}^{;\alpha}&-&2f_{(\mu; \nu)} + \hat{g}_{\mu\nu}f_{\alpha}{}^{;\alpha}-2\psi_{\alpha\beta}\hat{R}^{\alpha}{}_{\mu\nu}{}^\beta +2\psi_{\alpha(\mu}R^{\alpha}{}_{\nu)}\nonumber \\ &-&h_{\mu\nu}\hat{R}+\hat{g}_{\mu\nu}h_{\alpha\beta}\hat{R}^{\alpha\beta}= -2\delta T_{\mu\nu}\,,
\end{eqnarray}
where we have defined $\psi_{\mu\nu}=h_{\mu\nu}-(1/2)\hat{g}_{\mu\nu}(\hat{g}^{\alpha\beta}h_{\alpha\beta})$, and $f_\mu=\psi_{\mu\nu}{}^{;\nu}$. We also introduce the trace of this object, defined as $\psi=\psi_{\mu\nu}\hat{g}^{\mu\nu}$. Here, all covariant derivatives are taken with respect to the background metric. Note that under a linear coordinate transformation $x^{\mu}\rightarrow x^\mu+\xi^\mu$ the equations of motion are invariant, but the individual quantities previously defined change as:
\begin{equation}
	f_{\mu}\rightarrow f_{\mu}-\xi_\mu{}^{;\nu}{}_{;\nu}+\hat{R}_{\mu\nu}\xi^\nu; \quad \psi\rightarrow \psi +2 \xi^\alpha{}_{;\alpha}\,,
\end{equation}
where $\xi^\mu$ is an arbitrary small function of the space-time coordinates $x^\mu$.
Then, we can fix the gauge parameter $\xi^\mu$ to have $f_\mu=\psi=0$. Note that this gauge choice corresponds to the Lorentz gauge where $\psi_{\mu\nu}=h_{\mu\nu}$, $h=0$, and
\begin{equation}
\nabla^{\mu} h_{\mu\nu}=0\,.\label{LorentzEq}
\end{equation}
We see that eq.~$(\ref{LorentzEq})$ and the traceless condition give five constraints of the ten degrees of freedom of the metric perturbation. As we will confirm later, there will be a residual gauge freedom to be fixed in order to reduce the system to the two physical degrees of freedom of the massless graviton. The residual freedom can be explicitly identified and fixed once the background $\hat{g}_{\mu\nu}$ is chosen. 

From now on we assume that we are outside the lens, and hence in vacuum where the stress energy tensor vanishes for the background and perturbations. In this case, the background metric is a solution to the Einstein equations in vacuum. In the Lorentz gauge, the equations of motion then become: 
\begin{equation}
	h_{\mu\nu; \alpha}{}^{;\alpha}-2h_{\alpha\beta}\hat{R}^{\alpha}{}_{\mu\nu}{}^\beta + 2h_{\alpha(\mu}R^{\alpha}{}_{\nu)}-h_{\mu\nu}\hat{R}+\hat{g}_{\mu\nu}h_{\alpha\beta}\hat{R}^{\alpha\beta}= 0\,,
\end{equation}
and using the background equations for $\hat{g}_{\mu\nu}$, one is left with
\begin{equation}\label{GWeq}
	h_{\mu\nu; \alpha}{}^{;\alpha}-2h_{\alpha\beta}\hat{R}^{\alpha}{}_{\mu\nu}{}^\beta =0\,.
\end{equation}

In curved spacetimes, eq.\,(\ref{GWeq}) cannot be solved explicitly, except in cases of high symmetry. Furthermore, on a general curved background, we cannot define exact plane waves. However, in many cases of interest for gravitational lensing, one is interested on waves which appear as nearly plane on a scale large with respect to a typical wavelength of the wave, but small compared to the radius of curvature of the curved background on which the wave propagates. In analogy to these locally plane waves, we parametrize  $h_{\mu\nu}$ in the following form:
\begin{equation}\label{ansatz}
	h_{\mu\nu}= \Re(\epsilon_{\mu\nu}e^{i\Phi})\,,
\end{equation}
where $\Re$ is the real part of the expression in parenthesis.  
Here, $\Phi(x)$ is a real scalar function of the coordinates describing the phase of the waves, whereas $\epsilon_{\mu\nu}(x)$ is a symmetric complex tensor describing the polarization and amplitude. The parametrization (\ref{ansatz}) will become intuitive in the limit of geometric optics, where there is a natural split into a fast varying phase, $\Phi$, that describes a large wave frequency, and a slowly varying part, $\epsilon_{\mu\nu}$, that describes a smoothly evolving wave amplitude. Next, we insert this Ansatz into the Einstein equations:
\be\label{EE0}
\hat{\Box} \Re(\epsilon_{\mu\nu}e^{i\Phi})-2\hat{R}^{\alpha}{}_{\mu\nu}{}^\beta  \Re(\epsilon_{\alpha\beta}e^{i\Phi})=0\,,
\ee
which can be explicitly written as
\begin{eqnarray}\label{EE}
-k_\beta k^\beta \epsilon_{\mu\nu}&+&i [\textcolor{black}{2} k_\beta \epsilon_{\mu\nu}{}^{;\beta}+k^\beta{}_{;\beta} \epsilon_{\mu\nu}] \nonumber\\&+&\epsilon_{\mu\nu;\alpha}{}^{;\alpha}-2\epsilon_{\alpha\beta}\hat{R}^\alpha {}_{\mu\nu}{}^{\beta}  =0\,, 
\end{eqnarray}
where we have defined $k_\beta=\Phi_{;\beta}$ as the gradient of the phase.  Similarly, the gauge eq.~(\ref{LorentzEq}) gives:
\be\label{gaugeeq}
\nabla^{\mu} \epsilon_{\mu\nu}=-i k^{\mu} \epsilon_{\mu\nu} \,,
\ee
together with the traceless choice $\epsilon^{\mu}{}_{\mu}=0$. 

In geometric optics, the last two terms of eq.~(\ref{EE}) are systematically neglected, which is legitimate only in the weak field regime where the curvature contribution becomes small. More precisely, as discussed in \cite{Takahashi:2005sxa,Takahashi:2005ug} the weak-field approximation is valid when $(r_E/b)^2\ll 1$, where $b$ is the impact parameter, otherwise non-linearities become large. However, even in the weak field regime one can get wave effects if the wavelength is sufficiently large. In general, we expect the size of beyond geometric optics corrections to be of order 
 $\epsilon^{(1)}/A_I\sim (\lambda/r_S)(r_S/b)(r_E/b)^2$ (with respect to the incident amplitude $A_I$ of the wave), which will be small when the wavelength is small compared to the Schwarzschild radius of the lens, and when the impact parameter is much larger than $r_E$ (and $r_S$), i.e.~$\lambda \ll r_S\ll b$. Since the typical size of weak lensing effects in geometric optics is of order $\epsilon^{(0)}_{\text{lens}}\sim(r_E/b)^2$, then the relative suppression of wave corrections is expected to be of order $\epsilon^{(1)}/\epsilon^{(0)}_{\text{lens}}\sim (\lambda/b)$.
 
In order to illustrate this, let us consider the case of a wave traveling past a point-like lens described by a  Schwarzschild metric, with radius $r_S$. In this case, the gravitational potential is only a function of the radial coordinate $r$, $\phi(r)\sim r_S/r$ which we will assume to be much smaller than one. Then, we estimate the contributions of the last two terms in  eq.~(\ref{EE}) as (at linear order in the metric potential $\phi$): 
\be\label{111}
 \partial^2\epsilon +  \epsilon R \sim \epsilon \phi/r^2\,,
\ee
while the terms in square bracket in eq. (\ref{EE}) give:
\be\label{22}
(\partial k \epsilon+ k \partial \epsilon) \sim  \frac{\partial k}{k} k\epsilon\sim \epsilon \phi/(r\lambda)   \,,
\ee
where we used $\partial k/k\sim \phi/r$. It follows that for geometric optics to be valid one needs the terms in eq.~(\ref{22}) to be much smaller than those in (\ref{111}), i.e.~$\lambda\ll b$.


In this regime, we compute beyond geometric optics corrections using a perturbative approach. We introduce a large dimensionless parameter $\omega$ and expand the phase and tensor of eq.~(\ref{ansatz}) in the following way:
\begin{align}\label{defomega}
&\Phi \rightarrow \omega \Phi\,,\\ 
&\epsilon_{\mu\nu}\rightarrow \epsilon_{\mu\nu}^{(0)}+ \omega^{-1}\epsilon_{\mu\nu}^{(1)}+ \omega^{-2}\epsilon_{\mu\nu}^{(2)}+\dots\,. \label{EpsilonExpansion}
\end{align}
This expansion corresponds to a beyond-WKB approximation for a spin-2 wave. The parameter $\omega$ is introduced for book-keeping  but at the end one can absorb and ignore it. Also, $\Phi$ and $ \epsilon_{\mu\nu}$ are assumed to not depend on $\omega$. The limit $\omega\rightarrow \infty$ will describe the geometric optics limit, in which the phase $\omega\Phi$ changes rapidly compared to the amplitude of the wave, and whose gradient will describe the momentum of the geometric optics wave.
Note that we could have also expanded the phase in powers of $\omega$, and the result would have been equivalent to (\ref{ansatz}) with rescaled and shifted amplitudes $\epsilon_{\mu\nu}^{(n)}$\footnote{However, if we did so we would have a superposition of multiple waves with different momenta and the total perturbed phase of the wave would not have a clear physical interpretation anymore. For this reason, the only case in which the phase has a direct physical meaning is in the geometric optics regime, which motivates our choice of expanding only the complex amplitude $\epsilon_{\mu\nu}$ in eq.~(\ref{EpsilonExpansion}).}.

In what follows, we write down explicitly the equations of motion for the leading and subleading terms in the expansion (\ref{EpsilonExpansion}) in order to describe the geometric optics regime and its corrections.

\subsection{Geometric Optics}
At leading and next-to-leading order in $\omega$ we have the following equation that describes the geometric optics regime:
\begin{equation}
-\omega^2 k_\beta k^\beta \epsilon^{(0)}_{\mu\nu}+i\omega [2 k_\beta \epsilon^{(0)}_{\mu\nu}{}^{;\beta}+k^\beta{}_{;\beta} \epsilon^{(0)}_{\mu\nu}]+\mathcal{O}(\omega^0)=0\,,
\end{equation}
which then leads to two separate conditions for each order in $\omega$:
\begin{align}
& k_\beta k^\beta=0\,,\label{go:null}\\
&2 k_\beta \epsilon^{(0)}_{\mu\nu}{}^{;\beta}+k^\beta{}_{;\beta} \epsilon^{(0)}_{\mu\nu}=0\,.\label{go:linear}
\end{align}
From eq.~(\ref{go:null}) we see that $k^\mu$ is a null vector and thus gravitational waves propagate at the speed of light. Since $k^\mu$ is also a gradient we have that it inevitably satisfies the null geodesic equation
\begin{equation}\label{geodesic}
k^\mu k_{\nu;\mu}=0\,.
\end{equation}
The gauge condition at leading order in $\omega$ gives 
\begin{equation}
k^\mu \epsilon^{(0)}_{\mu\nu}=0\,,
\end{equation}
which indicates that the polarization is a transverse tensor. We note that we can separate $\epsilon_{\mu\nu}$ into an amplitude and polarization part as:
 \be\label{AAA}
 \epsilon_{\mu\nu}^{(0)}=A  A_{\mu\nu}\,,
 \ee
 with $A=\sqrt{\epsilon^{0*}_{\mu\nu}\epsilon^{0\mu\nu}}$ and $A_{\mu\nu}A^{\mu\nu}=1$\,, and simplify eq.~(\ref{go:linear}) using the gauge condition to obtain: 
\begin{align}
& k^\mu A_{;\mu}=-\frac{1}{2}\theta A; \quad \theta = k^\mu{}_{;\mu}\label{conservation}\\
& k^\alpha A_{\mu\nu;\alpha}=0\,.
\end{align}
These equations indicate that polarization is parallel-propagated along the null vector $k^\mu$, and lead to the covariant conservation of flux (i.e.~$(A^2k^\alpha)_{;\alpha}=0$). Indeed, we can define a four-momentum of the gravitons as $P^{\mu}\equiv \hbar k^{\mu}$ and introduce $N^{\mu}=A^2/(\hbar^2) P^{\mu}$ that we interpret as the graviton number current density. Then, we have $\nabla_{\mu} N^{\mu}=0$ which implies (via the Gauss's theorem) that the number of gravitons in a ray bundle is conserved and observer-independent. From this point of view, we can treat an incoherent gravitational radiation field as a graviton gas whose state is given by a distribution function on phase space. 


\subsection{Beyond Geometric Optics}\label{beyond}

Next, we take into account the leading order corrections beyond geometric optics. In the equation of motion (\ref{EE}) we collect sub-leading order terms in $\omega$ and obtain:
\begin{align}\label{EEbgo}
[\textcolor{black}{2} k_\beta \epsilon^{(1)}_{\mu\nu}{}^{;\beta}+k^\beta{}_{;\beta} \epsilon^{(1)}_{\mu\nu}] =S_{\mu\nu}^{(0)}\,, 
\end{align}
and the gauge condition gives
\be\label{gaugeeq}
k^{\nu} \epsilon^{(1)}_{\mu\nu}=S^{g(0)}_{\mu}\,,
\ee
where we have introduced the source-like tensors
\begin{align}\label{source1}
S_{\mu\nu}^{(0)}&= -i\left[2\epsilon^{(0)}_{\alpha\beta}\hat{R}^{\alpha}{}_{\mu\nu}{}^\beta - \epsilon^{(0)}_{\mu\nu}{}_{;\alpha}{}^{;\alpha}+(k^\alpha k_\alpha)\epsilon^{(2)}_{\mu\nu}  \right]\,,\\
S^{g(0)}_{\mu} &=i \nabla^\nu \epsilon^{(0)}_{\mu\nu}\,,\label{source2}
\end{align}
where the last term in eq.~(\ref{source1}) is actually vanishing since $k^{\mu}$ is a null vector. 
We see that the gauge equation (\ref{gaugeeq}) tells us that the polarization tensor beyond geometric optics is not transverse. The Einstein equation for the polarization tensor (\ref{EEbgo}) has now a source term and does not have a direct interpretation in terms of conservation of graviton flux. Note that these the two sources (\ref{source1})-(\ref{source2}) are the sole responsible for deviations beyond geometric optics. 

We emphasize that this procedure can be generalized at a generic order $n>0$ beyond geometric optics. In general, the Einstein and gauge equations will be given by:
\begin{align}\label{EEbgoN}
[\textcolor{black}{2} k_\beta \epsilon^{(n)}_{\mu\nu}{}^{;\beta}+k^\beta{}_{;\beta} \epsilon^{(n)}_{\mu\nu}] &=S^{(n-1)}_{\mu\nu}\,, \\
k^{\nu} \epsilon^{(n)}_{\mu\nu}&=S^{g(n-1)}_{\mu}\,,\label{Gaugen}
\end{align}
where
\begin{align}\label{sourcen}
S_{\mu\nu}^{(n-1)}&= -i\left[2\epsilon^{(n-1)}_{\alpha\beta}\hat{R}^{\alpha}{}_{\mu\nu}{}^\beta - \epsilon^{(n-1)}_{\mu\nu}{}_{;\alpha}{}^{;\alpha}  \right]\,,\nonumber\\
S^{g(n-1)}_{\mu} &=i \nabla^\nu \epsilon^{(n-1)}_{\mu\nu}\,.
\end{align}
The goal is then to construct an algorithm for solving eqs. (\ref{EEbgo}) with gauge (\ref{gaugeeq}) in order to reconstruct the total GW tensor including corrections beyond geometric optics, that is,
\begin{equation}\label{11}
h_{\mu\nu}= h_{\mu\nu}^{(0)} +  h_{\mu\nu}^{(1)}\,,
\end{equation}
with 
\begin{align}\label{12}
h_{\mu\nu}^{(0)}=\Re\{\epsilon^{(0)}_{\mu\nu}e^{i\omega\Phi }\}\,,\qquad h_{\mu\nu}^{(1)}= \Re\{\omega^{-1}\epsilon^{(1)}_{\mu\nu}e^{i\omega\Phi }\}\,.
\end{align}
In general, beyond geometric optics terms will add a correction to the phase and the amplitude of the wave, which, from eq.~(\ref{Gaugen}), will no longer be transverse to the wave vector $k^\mu$.This will generically be the case due to the fact that the wave will be diffracted, 
leading thus to a new total wave with oscillations in different directions. In this case, the polarization plane transverse to $k^\mu$ will be smeared and, as we will show in Section \ref{polarization}, new effective polarizations will arise.  

In the next section, we will discuss how to solve the equations of motion for $\epsilon^{(n)}_{\mu\nu}$.
However, we first mention that in order to find the solutions of this system, we will have to fix the  residual gauge freedom. Indeed, eq.~(\ref{Gaugen}) and the traceless condition, at each given perturbative order provide five algebraic conditions on $\epsilon_{\mu\nu}^{(n)}$. This implies that there are three residual gauge parameters that remain to be fixed. We will explicitly discuss how to fix the gauge in a specific example in Section \ref{computation}.

We conclude this section with a remark. We observe that the real and the imaginary part of $\epsilon_{\mu\nu}^{(2n-1)}$ source the imaginary and the real part of $\epsilon_{\mu\nu}^{(2n)}$, respectively. As an example, if the geometric optics polarization tensor $\epsilon_{\mu\nu}^{(0)}$ is real (as would be the case for a linearly polarized wave from a black hole binary edge-on along the line of sight), then $\epsilon_{\mu\nu}^{(2n)}$ will be real and $\epsilon_{\mu\nu}^{(2n-1)}$ will be purely imaginary. In this case, we can rewrite the first order correction in eq.\,(\ref{12}) using the fact that $\epsilon_{\mu\nu}^{(1)}$ is purely imaginary, as
\be\label{linear}
h^{(1)}_{\mu\nu}=\Re\left\{|\epsilon_{\mu\nu}^{(1)}|e^{i(\Phi+\pi/2)}\right\}\,. 
\ee
Hence we see that the total leading correction beyond geometric optics is a wave which has an additional contribution of  $\pi/2$ to the phase.


\section{Computational Technique}
\label{computation}

 In this section, we propose an algorithm to compute beyond geometric optics corrections in a recursive way in terms of decoupled first-order differential equations. We introduce a tetrad of null vectors (parallel transported along the geodesic associated to $k^{\mu}$)\footnote{Note that we choose the tetrad such that the geodesic equation is also satisfied for $m^\mu$ and $\ell^\mu$, which in turn means that it will also be satisfied for $n^\mu$. Therefore, all the properties of the tetrad are maintained as the wave propagates in the geometric optics limit. }
\be\label{tetrad}
\{k^{\mu}\,, m^{\mu}\,, \ell^{\mu}\,,n^{\mu}\}\,,
\ee
where $n^{\mu}$ is real, and $m^{\mu}$ and $\ell^{\mu}$ are complex such that:
\be\label{tetradeqns}
\ell^{\mu *}=m^{\mu}\,,\quad g_{\mu\nu}m^{\mu} \ell^{\nu}=1\,,\quad g_{\mu\nu}k^{\mu} n^{\nu}=-1\,, 
\ee
with all other contractions vanishing. This null tetrad forms a complete basis for real 4-vectors, and the spacetime metric can always be written as:
\begin{equation}
g_{\mu\nu}=m_\mu \ell_\nu + m_\nu\ell_\mu -n_\mu k_\nu - n_\nu k_\mu\,.
\end{equation}
Next, we use the null tetrad to form a basis for the rank-2 tensor $\epsilon_{\mu\nu}$ as:
\begin{align}\label{decomposition}
\epsilon_{\mu\nu}=&\alpha_{nk} \Theta_{\mu\nu}^{nk}+\alpha_{m\ell} \Theta_{\mu\nu}^{m\ell}+\alpha_{nn} \Theta_{\mu\nu}^{nn}+\alpha_{kk} \Theta_{\mu\nu}^{kk}+\alpha_{nm}\Theta_{\mu\nu}^{nm}\nonumber \\
&+\alpha_{n\ell}\Theta_{\mu\nu}^{n\ell}+\alpha_{km}\Theta_{\mu\nu}^{km}+\alpha_{k\ell}\Theta_{\mu\nu}^{k\ell}+\alpha_{mm}\Theta_{\mu\nu}^{mm}+\alpha_{\ell\ell}\Theta_{\mu\nu}^{\ell\ell} \,,
\end{align}
where $\alpha_{AB}$ are all complex coefficients whereas $\Theta_{\mu\nu}$ are operators constructed from the null tetrad and form a basis for rank-2 tensors. Explicitly, there will be 10 symmetric operators defined as:
\be\label{baseop}
\Theta_{\mu\nu}^{AB}\equiv \frac{1}{2}\left(A_{\mu}B_{\nu}+A_{\nu}B_{\mu}\right)\,,
\ee
with $(A, B)=\{k, m, \ell, n\}$. For future reference, we mention properties that these operators satisfy due to the orthogonality of the null tetrad:
\begin{align}
(\Theta_{m\ell})_{\mu\nu}(\Theta_{m\ell})^{\mu\nu}&=\frac{1}{2}\,,\label{ortoI}\\
(\Theta_{nk})_{\mu\nu}(\Theta_{nk})^{\mu\nu}&=\frac{1}{2}\,,\\
(\Theta_{km})_{\mu\nu}(\Theta_{n\ell})^{\mu\nu}&=-\frac{1}{2}\,,\\
(\Theta_{k\ell})_{\mu\nu}(\Theta_{nm})^{\mu\nu}&=-\frac{1}{2}\,,\\
(\Theta_{kk})_{\mu\nu}(\Theta_{nn})^{\mu\nu}&=1\,,\\
(\Theta_{mm})_{\mu\nu}(\Theta_{\ell\ell})^{\mu\nu}&=1\,,\label{ortoF}
\end{align} 
with all other scalar contractions vanishing. 
We use this decomposition to obtain a set of equations of motion for the coefficients $\alpha_{AB}$ such that the Einstein and gauge equations are satisfied. Explicitly, we replace eq.~(\ref{decomposition}) into (\ref{EE}) and obtain:
\begin{align}\label{eqP}
2 \mathcal{D} \alpha_{nk}+\nabla_{\alpha}k^{\alpha}\alpha_{nk}&=2 S_{\mu\nu}(\Theta_{nk})^{\mu\nu}\,,\\
2 \mathcal{D} \alpha_{m\ell}+\nabla_{\alpha}k^{\alpha}\alpha_{m\ell}&=2 S_{\mu\nu}(\Theta_{m\ell})^{\mu\nu}\,,\\
2 \mathcal{D} \alpha_{kk}+\nabla_{\alpha}k^{\alpha}\alpha_{kk}&= S_{\mu\nu}(\Theta_{nn})^{\mu\nu}\,,\\
2 \mathcal{D} \alpha_{nn}+\nabla_{\alpha}k^{\alpha}\alpha_{nn}&= S_{\mu\nu}(\Theta_{kk})^{\mu\nu}\,,\\
2 \mathcal{D} \alpha_{nm}+\nabla_{\alpha}k^{\alpha}\alpha_{nm}&=2 S_{\mu\nu}(\Theta_{k\ell})^{\mu\nu}\,,\\
2 \mathcal{D} \alpha_{n\ell}+\nabla_{\alpha}k^{\alpha}\alpha_{n\ell}&=2 S_{\mu\nu}(\Theta_{km})^{\mu\nu}\,,\\
2 \mathcal{D} \alpha_{km}+\nabla_{\alpha}k^{\alpha}\alpha_{km}&=2 S_{\mu\nu}(\Theta_{n\ell})^{\mu\nu}\,,\\
2 \mathcal{D} \alpha_{k\ell}+\nabla_{\alpha}k^{\alpha}\alpha_{k\ell}&=2 S_{\mu\nu}(\Theta_{nm})^{\mu\nu}\,,\\
2 \mathcal{D} \alpha_{mm}+\nabla_{\alpha}k^{\alpha}\alpha_{mm}&= S_{\mu\nu}(\Theta_{\ell \ell})^{\mu\nu}\,,\\
2 \mathcal{D} \alpha_{\ell\ell}+\nabla_{\alpha}k^{\alpha}\alpha_{\ell\ell}&= S_{\mu\nu}(\Theta_{mm})^{\mu\nu}\,,
\end{align}
where we have introduced the directional derivative along $k^{\mu}$ defined as
\be
\mathcal{D}\equiv k^{\mu}\nabla_{\mu}=\frac{D}{d\lambda}\,,
\ee
where $\lambda$ is an affine parameter along the graviton geodesic. Similarly, the Lorentz gauge equation gives:
\begin{align}\label{gauge1P}
\alpha_{nk}&=2S^g_{\mu}n^{\mu}\,,\\
\alpha_{nn}&=S^g_{\mu}k^{\mu}\,,\\
\alpha_{n m}&=-2 S^g_{\mu}\ell^{\mu}\,,\\
\alpha_{n \ell}&=-2S^g_{\mu}m^{\mu}\,.
\end{align}
The gauge condition on the trace is equivalent to $g^{\mu\nu}\epsilon_{\mu\nu}=0$ and gives the condition 
\be\label{gauge2P}
\alpha_{nk}=\alpha_{m\ell}\,.
\ee

We can use this formalism to revisit the geometric optics limit and its corrections. In the geometric optics limit, all the source terms in the equations for the coefficients $\alpha_{AB}$ vanish. Therefore, from the gauge conditions (\ref{gauge1P})-(\ref{gauge2P}) it follows that in this regime the following coefficients will vanish 
\begin{equation}
\alpha^{(0)}_{nk}=\alpha^{(0)}_{nn}=\alpha^{(0)}_{nm}=\alpha^{(0)}_{n\ell}=\alpha^{(0)}_{m\ell}=0\,,
\end{equation}
whereas the remaining five coefficients $\left\{ \alpha^{(0)}_{kk}, \alpha^{(0)}_{km}, \alpha^{(0)}_{k\ell}, \alpha^{(0)}_{mm}, \alpha^{(0)}_{\ell\ell}\right\}$ will satisfy the same schematic equation:
\begin{equation}
2 \mathcal{D} \alpha_{AB}^{(0)} + \nabla_{\alpha}k^{\alpha}\alpha_{AB}^{(0)}= 0\,.
\end{equation}

Next, we expand the Einstein and gauge equations to obtain the leading order beyond geometric optics (note that equations for higher order corrections will have the same structure). The source terms are given by:
\begin{align}\label{SS1}
&S_{\mu\nu}^{(0)}=-i2\epsilon^{(0)}_{\alpha\beta}\hat{R}^\alpha {}_{\mu\nu}{}^{\beta}+ i\epsilon^{(0)}_{\mu\nu;\alpha}{}^{;\alpha}\,,\\
&S_{\mu\nu}^{g(0)}= i \nabla^{\mu}\left(\epsilon_{\mu\nu}^{(0)}\right)\,, \label{SS2}
\end{align}
and the equations of motion become:
\begin{align}
2\mathcal{D}\alpha_{nk}^{(1)}+\nabla_\alpha k^\alpha \alpha_{nk}^{(1)}&=2S_{\mu\nu}^{(0)}(\Theta_{nk})^{\mu\nu}\,,\label{Ank1}\\
2 \mathcal{D} \alpha_{m\ell}^{(1)} + \nabla_{\alpha}k^{\alpha}\alpha_{m\ell}^{(1)} &=2 S_{\mu\nu}^{(0)}(\Theta_{m\ell})^{\mu\nu}\,,\label{Aml1}\\
2 \mathcal{D} \alpha_{kk}^{(1)}+\nabla_{\alpha}k^{\alpha}\alpha_{kk}^{(1)}&= S_{\mu\nu}^{(0)}(\Theta_{nn})^{\mu\nu}\,,\label{Akk1}\\
2 \mathcal{D} \alpha_{nn}^{(1)}+\nabla_{\alpha}k^{\alpha}\alpha_{nn}^{(1)}&= S_{\mu\nu}^{(0)}(\Theta_{kk})^{\mu\nu}\,,\label{Ann1}\\
2 \mathcal{D} \alpha_{nm}^{(1)}+\nabla_{\alpha}k^{\alpha}\alpha_{nm}^{(1)}&=2 S_{\mu\nu}^{(0)}(\Theta_{k\ell})^{\mu\nu}\,,\label{Anm1}\\
2 \mathcal{D} \alpha_{n\ell}^{(1)}+\nabla_{\alpha}k^{\alpha}\alpha_{n\ell}^{(1)}&= 2S_{\mu\nu}^{(0)}(\Theta_{km})^{\mu\nu}\,,\label{Anl1}\\
2 \mathcal{D} \alpha_{km}^{(1)}+\nabla_{\alpha}k^{\alpha}\alpha_{km}^{(1)}&=2 S_{\mu\nu}^{(0)}(\Theta_{n\ell})^{\mu\nu}\,,\\
2 \mathcal{D} \alpha_{k\ell}^{(1)}+\nabla_{\alpha}k^{\alpha}\alpha_{k\ell}^{(1)}&= 2S_{\mu\nu}^{(0)}(\Theta_{nm})^{\mu\nu}\,,\\
2 \mathcal{D} \alpha_{mm}^{(1)}+\nabla_{\alpha}k^{\alpha}\alpha_{mm}^{(1)}&= S_{\mu\nu}^{(0)}(\Theta_{\ell \ell})^{\mu\nu}\,,\label{Amm1}\\
2 \mathcal{D} \alpha_{\ell\ell}^{(1)}+\nabla_{\alpha}k^{\alpha}\alpha_{\ell\ell}^{(1)}&= S_{\mu\nu}^{(0)}(\Theta_{mm})^{\mu\nu}\,, \label{All1}
\end{align}
where all the operators on the right hand side are evaluated in the geometric optics limit. Similarly, the gauge equation gives:
\begin{align}
\alpha_{nk}^{(1)}&=2S^{g(0)}_{\mu}n^{\mu}\,,\\
\alpha_{nn}^{(1)}&=S^{g(0)}_{\mu}k^{\mu}\,,\label{Alpha1nn}\\
\alpha_{n m}^{(1)}&=-2 S^{g(0)}_{\mu}\ell^{\mu}\,,\\
\alpha_{n \ell}^{(1)}&=-2S^{g(0)}_{\mu}m^{\mu}\,,
\end{align}
and the traceless condition gives
\be
\alpha_{nk}^{(1)}=\alpha_{m\ell}^{(1)}\,.
\ee
We see that beyond geometric optics, in principle all the components of the polarization tensor are sourced and thus non vanishing. We also note that the gauge conditions fix some of the same coefficients that satisfy also eq.~(\ref{Ank1}), (\ref{Aml1}), (\ref{Ann1}), (\ref{Anm1}), (\ref{Anl1}). This means that in these cases there will be relations between the sources $S^{(0)}_{\mu\nu}(\Theta_{AB})^{\mu\nu}$ and $S^{(0)}_\mu A^\mu$, which will be automatically satisfied by construction.

As we have mentioned before, in order to find the solution to these equations, we must fully fix the gauge. We assume that sufficiently far from the source and the lens, the spacetime is nearly flat. Then far from the source we have still the freedom to transform the (total) polarization tensor as 
\be\label{residual}
\epsilon_{\mu\nu}\rightarrow \epsilon_{\mu\nu}+C_{\mu}k_{\nu}+C_{\nu}k_{\mu}\,,
\ee
where $C_{\mu}$ is a complex arbitrary vector orthogonal to $k_{\mu}$. This gauge transformation preserves both the Lorentz gauge and traceless condition in Minkowski. 
We fix this freedom in the geometric optics limit by imposing that $n^{\mu}\epsilon^{(0)}_{\mu\nu}=0$ near emission. 
As a consequence, we  end up with the following additional coefficients vanishing:
\begin{equation}
 \alpha^{(0)}_{kk}= \alpha^{(0)}_{km}= \alpha^{(0)}_{k\ell}=0\,.
\end{equation}
We  are therefore left with only two non-zero amplitudes:
 \begin{equation}\label{Epsilon0}
\epsilon^{(0)}_{\mu\nu}= \alpha^{(0)}_{mm} m_\mu m_\nu +  \alpha^{(0)}_{\ell\ell} \ell_\mu \ell_\nu\,.
\end{equation}
These amplitudes (which are complex) require four initial conditions to be fully fixed, and describe the two possible polarizations of gravitational waves for a massless graviton. 

For the corrections beyond geometric optics, since the source terms (\ref{SS1}) and (\ref{SS2}) are vanishing on a flat background, a natural gauge choice is to set all the coefficients $\alpha^{(n)}_{AB}$ beyond geometric to zero near emission. Hence in that region we have $\epsilon_{\mu\nu}=\epsilon_{\mu\nu}^{(0)}$. This choice fixes the initial conditions for all the first-order differential equations  for $\alpha^{(n)}_{AB}$. 
As the wave propagates, corrections to $(\ref{Epsilon0})$ are generated, sourced by  (\ref{SS1}) and (\ref{SS2}) which are non-zero on a curved background. However, these are not new degrees of freedom of the wave but they rather represent additional effects that appear from taking the curvature of the background into consideration. 


\subsection{An example: geometric optics for point-like lens}
In order to illustrate the use of the technique proposed, we solve the geometric optics order in an explicit situation: GW lensed by a point-like lens in the weak field regime. Let us then consider the following spherically symmetric background spacetime:
\be
ds^2=\bar{g}_{\mu\nu}dx^{\mu}dx^{\nu}=-(1+2\phi)dt^2+(1-2\phi)d\Omega^2\,,
\ee
with $d\Omega^2=r^2(d\theta^2+\sin\theta^2 d\varphi^2)$ and $\phi=\phi(\vec{x})$ is the gravitational potential of the lens which we assume to be weak outside the lens,  $\phi\ll1$. This spacetime will be valid for scenarios where we have a massive astrophysical object acting as a lens and we study the behaviour of gravitational waves far enough from the object. 

Outside the source and sufficiently far from the lens, the spacetime is flat so we can solve the equations for the geometric optics regime perturbatively around Minkowski. We thus introduce the Minkowski tetrad $\{\bar{k}^\mu, \bar{m}^\mu, \bar{\ell}^\mu, \bar{n}^\mu\}$. Explicitly, the wavevector $\bar{k}^\mu$ can be written as 
\begin{equation}
\bar{k}=(\bar{k}^0\,, \bar{k}^i)=E(1, -{\bf{e}})\,,
\end{equation}
where ${\bf{e}}$ is the direction of the spatial momentum normalized to unity: $|{\bf{e}}|^2=1$, and $E$ is a constant amplitude for the 4-vector momentum. Similarly, the rest of the tetrad can be chosen as 
\begin{equation}
\bar{n}^\mu= \frac{1}{2E}(1,{\bf{e}}),\quad \bar{m}^\mu=\frac{1}{\sqrt{2}}(0, {\bf{e}}_1+i{\bf{e}}_2)\,,
\end{equation} 
where ${\bf{e}}_{1,2}$ are real orthonormal 3D vectors, i.e.~$|{\bf{e}}_{1,2}|^2=1$ and ${\bf{e}}_1\cdot {\bf{e}}_2=0$, orthogonal to ${\bf{e}}$ as well, that is, ${\bf{e}}_{1, 2}\cdot{\bf{e}}=0$. For instance, for a wave traveling along $-z$ we would have:
\begin{equation}
\bar{k}^\mu=E(1,0,0,-1), \, \bar{n}^\mu=\frac{1}{2E}(1,0,0,1), \, \bar{m}^\mu=\frac{1}{\sqrt{2}}(0,1,i,0)\,.
\end{equation}
In Minkowski, the two coefficients $\bar{\alpha}_{mm}$ and $\bar{\alpha}_{\ell\ell}$ are constants that will be determined by initial conditions, which in turn describe different physical set ups (e.g.~the choice of $\bar{\alpha}_{mm}=1$ and $\bar{\alpha}_{\ell\ell}=0$ describes an emitted right-handed circularly-polarized wave).

Next, we calculate how the emitted plane wave gets modified when it propagates on a spacetime which is curved. We work in the weak field limit, and we reconstruct the wave at the observer at linear order in the metric potential $\phi$.
In particular, we expand the geometric optics vector $k^\mu$ as 
\be
k^{\mu}=\bar{k}^{\mu}+\delta k^{\mu}\,,
\ee 
where $\bar{k}^\mu$ is the 4-vector in Minkowski, and $\delta k^\mu$ is a small perturbation satisfying the linearized geodesic equation, which gives \cite{Cusin:2017fwz}:
\begin{align}
\delta k^{\mu}&=(\delta k^0\,, \delta k^i)=\left(-2 \bar{k}^0 \phi|_{\lambda_S}^{\lambda}\,, 2 \bar{k}^i \phi |_{\lambda_S}^{\lambda}-2 \int_{\lambda_S}^{\lambda} d\lambda E^2 \partial^i\phi\right)\,,
\end{align}
where we chose the affine parameter such that $d\lambda=-E^{-1}dz$ and $\lambda_S$ is the value of the affine parameter at the source. 
Similarly, we can use the geodesic equation to obtain $\delta m^\mu$:
\begin{align}
\delta m^\mu&= (\delta m^0, \delta m^i)\nonumber\\
&=\left(-\bar{k}^0\bar{m}^j\int_{\lambda_S}^{\lambda} d\lambda\, \partial_j\phi \,, \bar{m}^i\phi|_{\lambda_S}^{\lambda}+ \bar{k}^i\bar{m}^j\int_{\lambda_S}^{\lambda} d\lambda\, \partial_j\phi  \right)\,,
\end{align}
with $\delta \ell^\mu$ given by its complex conjugate.
Note that this solution indeed satisfies the orthogonality condition $k^\mu m_{\mu}$ to linear order, that is, $\delta k^\mu \bar{m}_\mu= -\delta m_\mu \bar{k}^\mu$. Using the rest of the properties of the tetrads we solve for $\delta n^{\mu}$ and obtain:
\begin{equation}
\delta n^\mu=-2\bar{n}^\mu\left( -\phi  |_{\lambda_S}^{\lambda} + \bar{n}^i  \int_{\lambda_S}^{\lambda} d\lambda E^2 \partial_i\phi\right)\,,
\end{equation}
which satisfies the relations $\delta n^\mu \bar{m}_{\mu}=\delta n^\mu \bar{n}_{\mu} =0$ and $\delta n^\mu \bar{k}_{\mu}=-\delta k_\mu \bar{n}^{\mu} $. 
Finally, we solve for the two coefficients $\alpha^{(0)}_{mm}$ and $\alpha^{(0)}_{\ell\ell}$:
\begin{align}\label{alpha}
&\delta\alpha_{mm}(\lambda)=-\frac{1}{2}\bar{\alpha}_{mm}\int_{\lambda_S}^{\lambda} d\lambda \nabla_{\alpha}k^{\alpha}\,,\\
&\delta\alpha_{\ell\ell}(\lambda)=-\frac{1}{2}\bar{\alpha}_{\ell\ell}\int_{\lambda_S}^{\lambda} d\lambda \nabla_{\alpha}k^{\alpha}\,,
\end{align}
which vanish at linear order in the potential.  It follows that the polarization tensor at any position $x\equiv x(\lambda)$, in geometric optics is given by
\begin{align}\label{tons}
\epsilon_{\mu\nu}^{(0)}(x)&=\bar{\alpha}_{mm}\Theta^{mm}_{\mu\nu}(x)+\bar{\alpha}_{\ell\ell}\Theta^{\ell\ell}_{\mu\nu}(x)\,, 
\end{align}
where the tensors $\Theta^{mm}_{\mu\nu}$ and $\Theta^{\ell\ell}_{\mu\nu}$ are built using $m^{\mu}=\bar{m}^{\mu}+\delta m^{\mu}$ and $\ell^{\mu}=\bar{\ell}^{\mu}+\delta \ell^{\mu}$ up to first order in perturbations. From this example we see that the total tensor $\epsilon^{(0)}$ will have a component coming from the Minkowski expansion in addition to a component generated exclusively by lensing $\epsilon^{(0)}_{\text{lens}}$, that is typically of the order of $\epsilon^{(0)}_{\text{lens}}/ \epsilon^{(0)}\sim (r_E/b)^2$. 
From this explicit example we see that the frequency of the wave does not change in geometric optics for far observers, and the only changes are given by the directions of the tetrad, which is parallel transported along the geodesic of gravitons. We also see that the amplitude of the tensor is not modified by the propagation. This is because in this specific example the right hand side of Eq. (\ref{conservation}) vanishes.  Finally, the ratio between left and right handed polarizations at the observer and at the source are the same. This is a consequence of the fact that the polarization tensor is parallel transported along a ray. 

\section{Polarization}
\label{polarization}
In a general metric theory, gravitational waves can have up to six different polarization modes corresponding to six independent degrees of freedom carried by the Riemann tensor. These components are encoded in the so-called Newman-Penrose (NP) scalars, which are given in terms of projections of the Weyl tensor of the wave on the null tetrad basis. Specifically, the six polarizations are encoded in the following quantities \cite{will_2018}
\begin{align}\label{NP1}
\Psi_2&=-\frac{1}{6} \mathcal{C}_{\mu\nu\alpha\beta}k^\mu n^\nu k^\alpha n^\beta\,,\\
\Psi_3&=  -\frac{1}{2}\mathcal{C}_{\mu\nu\alpha\beta}n^\mu k^\nu n^\alpha \ell^\beta\,,\\
\Psi_4&= -\mathcal{C}_{\mu\nu\alpha\beta}n^\mu \ell^\nu n^\alpha \ell^\beta\,,\\
\Phi_{22}&=  \mathcal{C}_{\mu\nu\alpha\beta}n^\mu m^\nu \ell^\alpha n^\beta\,,\label{NP2}
\end{align}
with all other projections being redundant or vanishing for this choice of tetrad. The scalars $\Psi_4$ and $\Psi_3$ are complex and describe helicity-2 and helicity-1 polarizations, respectively. The scalars $\Psi_2$ and $\Phi_{22}$ are real and describe spin-0 polarizations. Here, $\mathcal{C}_{\mu\nu\alpha\beta}$ is the Weyl tensor linear in the $h_{\mu\nu}$ perturbation. In this work, since we are considering perturbations in vacuum, the Weyl tensor is equal to the Riemann tensor. 

Next, we explicitly compute the Newman-Penrose scalars associated to a gravitational wave propagating on a vacuum solution, up to first order beyond geometric optics. Using doubled square brackets to denote independent anti-symmetrization over the inner and outer pairs of indices (for example $t_{[a[bc]d]}=\frac{1}{2}(t_{a[bc]d}-t_{d[bc]a})$ we can write the Riemann tensor as:
\be\label{general}
 \mathcal{R}_{\mu\nu\alpha\beta}=-2 \nabla_{[\mu} \nabla_{[\alpha}h_{\beta]\nu]}+R_{\mu\nu[\alpha}{}^{\gamma}h_{\beta]\gamma}\,.
 \ee
Replacing the expression for the metric (\ref{12}) and ordering powers of $\omega$ up to $\mathcal{O}(\omega^0)$, one obtains
\be
 \mathcal{R}_{\mu\nu\alpha\beta}= \mathcal{R}^{(0)}_{\mu\nu\alpha\beta}+\mathcal{R}^{(1)}_{\mu\nu\alpha\beta}\,,
 \ee
 where
 \begin{eqnarray}
\mathcal{R}^{(0)}_{\mu\nu\alpha\beta}=-2\omega^2 \Re\left\{e^{i\Phi}k_{[\mu}\epsilon^{(0)}_{\nu][\alpha}k_{\beta]}\right\}\,,
 \end{eqnarray}
  and
   \begin{eqnarray}\label{RR1}
&&\mathcal{R}^{(1)}_{\mu\nu\alpha\beta}=-2\omega \Re\{e^{i\Phi}k_{[\mu}\epsilon^{(1)}_{\nu][\alpha}k_{\beta]}\nonumber \\ &+&i e^{i\Phi}\left[(\nabla_{[\mu}\epsilon^{(0)}_{\nu][\alpha})k_{\beta]}+(\nabla_{[\alpha}\epsilon^{(0)}_{\beta][\mu})k_{\nu]}-(\nabla_{[\alpha}k_{[\mu})\epsilon^{(0)}_{\nu]\beta]}\right]\}\,.\nonumber \\
  \end{eqnarray}
We see that terms with the background Riemann in (\ref{general}) appear only two orders beyond geometric optics since they do not contain any derivative of the GW field. Since in geometric optics the polarization tensor is transverse, using the properties of the tetrad and in particular the fact that  $k_{\mu}n^{\nu}$ and $\ell_{\mu}m^{\mu}$ are the only non-vanishing contractions, it is straightforward to check that in geometric optics only $\Psi_4$ is non-vanishing.  We stress that the usefulness of the Newman-Penrose formalism resides on the fact that these are all gauge invariant variables. It follows that even if we did not fix completely the gauge at the level of $h_{\mu\nu}$, the Newman-Penrose scalars will contain only those components of $h_{\mu\nu}$ corresponding to physical degrees of freedom. 
  
In general, when using the geometric optics tetrad to project the Weyl and compute the Newman Penrose scalars beyond geometric optics, polarizations other then  $\Psi_4$ will arise, as recently discussed in Ref.\,\cite{Harte:2018wni} for both the case of electromagnetic and gravitational wave. We compute here their explicit expression in terms of the coefficients $\alpha_{AB}^{(1)}$ introduced in the previous section. 

All the Newman-Penrose scalars will have the same  schematic form:
 \begin{equation}
 p \mathcal{C}_{\mu\nu\alpha\beta}A^\mu B^\nu C^\alpha D^\beta, 
 \end{equation}
where $p$ denotes the numerical pre-factor in the definitions (\ref{NP1})-(\ref{NP2}), and the vectors $\{A^\mu,B^\mu,C^\mu,D^\mu\}$ correspond to specific vectors of the null tetrad. We can rewrite this general expression as:
 \begin{align}
 &2p \omega \cos(\omega\Phi) \left[-\omega \Re \left(r_{0\mu\nu\alpha\beta}\right)-  \Re\left( r_{2\mu\nu\alpha\beta}\right)\nonumber\right. \\
 & \left. + \Im (r_{1\mu\nu\alpha\beta})   \right] A^\mu B^\nu C^\alpha D^\beta \nonumber\\
 &+2p \omega\sin(\omega\Phi)\left[ \omega \Im \left(r_{0\mu\nu\alpha\beta}\right) + \Im \left( r_{2\mu\nu\alpha\beta}\right)\nonumber \right. \\
 & \left. +  \Re(r_{1\mu\nu\alpha\beta})  \right] A^\mu B^\nu C^\alpha D^\beta ,
 \end{align}
 where we have defined
 \begin{align}
 r_{0\mu\nu\alpha\beta}&\equiv k_{[\mu}\epsilon^{(0)}_{\nu][\alpha}k_{\beta]}, \\
 r_{1\mu\nu\alpha\beta}&\equiv \left[(\nabla_{[\mu}\epsilon^{(0)}_{\nu][\alpha})k_{\beta]}+(\nabla_{[\alpha}\epsilon^{(0)}_{\beta][\mu})k_{\nu]}-(\nabla_{[\alpha}k_{[\mu})\epsilon^{(0)}_{\nu]\beta]}\right],\\
 r_{2\mu\nu\alpha\beta}&\equiv k_{[\mu}\epsilon^{(1)}_{\nu][\alpha}k_{\beta]}.
 \end{align}
  Next, we assume that $\epsilon^{(0)}_{\mu\nu}$ is given by eq.~(\ref{tons}), whereas $\epsilon^{(1)}_{\mu\nu}$  is completely generic with all non-zero  $\alpha^{(1)}_{AB}$ coefficients. 
  In this case, we find the following expressions for the NP scalars:
  \begin{align}
 \Psi_{2}&=-\frac{1}{24}\omega \left[ \cos(\omega\Phi) \left(\alpha^{(1)}_{nn}+\alpha^{(1)*}_{nn} \right) + \nonumber \right. \\
 & \left. +i\sin(\omega\Phi) \left(\alpha^{(1)}_{nn}-\alpha^{(1)*}_{nn}\right) \right],
 \end{align} 
 which is manifestly real, and only contributes beyond geometric optics. Note that this can be rewritten solely in terms of the geometric optics solution by using eq.~(\ref{Alpha1nn}):
   \begin{align}
 \Psi_{2}&=\frac{1}{12}i\omega (\nabla^\mu k^\nu)\left[ \cos(\omega\Phi) \Re\left(\epsilon^{(0)}_{\mu\nu}\right) -\sin(\omega\Phi)  \Im\left(\epsilon^{(0)}_{\mu\nu}\right) \right],
 \end{align} 
We also obtain:
  \begin{align}
  \Psi_3&=  \frac{1}{16}\omega\cos(\omega\Phi)\left(\alpha^{(1)}_{nm}+ \alpha^{(1)*}_{n\ell}\right)\nonumber\\
  &+\frac{1}{16}i\omega\sin(\omega\Phi)\left(\alpha^{(1)}_{nm}- \alpha^{(1)*}_{n\ell}\right)\nonumber\\
  &+ \frac{1}{4}\omega m_\alpha n^\beta (\nabla_\beta k^\alpha) \left[ \sin(\omega\Phi)\left(  \alpha^{(0)}_{mm} + \alpha^{(0)*}_{\ell\ell}\right)\right.\nonumber\\
  & \left.  -i \cos(\omega\Phi)\left(  \alpha^{(0)}_{mm} - \alpha^{(0)*}_{\ell\ell}\right) \right] ,
  \end{align}
  which is a complex scalar, whose real and imaginary parts describe the two vector polarizations. Again, this can be fully rewritten in terms of the geometric optics solution using the gauge equations to obtain:
    \begin{align}
  \Psi_3&=  \frac{1}{4}\omega\cos(\omega\Phi)\ell^\mu\left(\nabla^\nu \Im(\epsilon^{(0)}_{\nu\mu})\right)\nonumber\\
  &+\frac{1}{4}\omega\sin(\omega\Phi))\ell^\mu\left(\nabla^\nu \Re(\epsilon^{(0)}_{\nu\mu})\right)\nonumber\\
  &+ \frac{1}{4}\omega m_\alpha n^\beta (\nabla_\beta k^\alpha) \left[ \sin(\omega\Phi)\left(  \alpha^{(0)}_{mm} + \alpha^{(0)*}_{\ell\ell}\right)\right.\nonumber\\
  & \left.  -i \cos(\omega\Phi)\left(  \alpha^{(0)}_{mm} - \alpha^{(0)*}_{\ell\ell}\right) \right].
  \end{align}
   In addition, we find:
 \begin{align}
\Psi_4&=- \frac{1}{4}\omega\cos(\omega\Phi)\left[\omega \left(\alpha^{(0)}_{mm}+\alpha^{(0)*}_{\ell\ell}\right)+\left(\alpha^{(1)}_{mm}+\alpha^{(1)*}_{\ell\ell}\right)\nonumber\right.\\
&\left.  - 8 n^\alpha \ell^\beta\ell^\mu\nabla_{[\alpha}\Im(\epsilon^{(0)}_{\beta]\mu})  - i \left(\alpha^{(0)}_{mm} - \alpha^{(0)*}_{\ell\ell} \right)n^\alpha n^\beta (\nabla_\beta k_\alpha)   \right] \nonumber\\
&-\frac{1}{4}i\omega\sin(\omega\Phi)\left[\omega \left(\alpha^{(0)}_{mm}-\alpha^{(0)*}_{\ell\ell}\right)+ \left(\alpha^{(1)}_{mm}-\alpha^{(1)*}_{\ell\ell}\right)\right. \nonumber\\
& \left.+8i n^\alpha \ell^\beta\ell^\mu\nabla_{[\alpha}\Re(\epsilon^{(0)}_{\beta]\mu})  - i \left(\alpha^{(0)}_{mm} + \alpha^{(0)*}_{\ell\ell} \right)n^\alpha n^\beta (\nabla_\beta k_\alpha) \right]\,.
 \end{align} 
From here we explicitly confirm that $\Psi_4$ is the only non-zero NP scalar in the geometric optics regime, whose real and complex components are determined by the amplitudes $\alpha_{mm}^{(0)}$ and $\alpha_{\ell\ell}^{(0)}$, which describe the two tensor polarizations as expected. 
Finally, we obtain the last NP scalar:
 \begin{align}
\Phi_{22}&=  \frac{1}{2}\omega\sin(\omega\Phi)\left[(\nabla_\alpha n^\beta) \left( \frac{1}{2}(\alpha^{(0)}_{mm} + \alpha^{(0)*}_{\ell\ell} )m^\beta m^\alpha + c.c. \right) \right. \nonumber\\
&\left.+ 2 m^\alpha \ell^\mu \left(n^\beta \nabla_\beta \Re(\epsilon^{(0)}_{\alpha\mu})\right) \right]\nonumber\\
&+  \frac{1}{2}\omega\cos(\omega\Phi)\left[(\nabla_\alpha n^\beta) \left(- \frac{i}{2}(\alpha^{(0)}_{mm} - \alpha^{(0)*}_{\ell\ell} )m^\beta m^\alpha + c.c. \right) \right. \nonumber\\
&\left.+ 2 m^\alpha \ell^\mu \left(n^\beta \nabla_\beta \Im(\epsilon^{(0)}_{\alpha\mu})\right) \right]\,,
 \end{align}
which is real as it can be explicitly verified using the fact that the polarization tensor is symmetric.

We conclude that all the NP scalars are fully determined by the components $\alpha_{mm}$ and $\alpha_{\ell\ell}$ at all orders, together with the null tetrad (which is fixed in the geometric optics regime). This means that from the full set of equations (\ref{Ank1})-(\ref{All1}) beyond geometric optics, we only need to solve (\ref{Amm1})-(\ref{All1}) for $\alpha_{mm}^{(1)}$ and $\alpha_{\ell\ell}^{(1)}$. All the other components are hence expected to be dependent or gauge artifacts. 

Note that the NP scalars depend on the chosen tetrad, and here we are projecting the Riemann tensor onto the parallel transported tetrad of weak lensing. In Appendix\,\ref{App:Tetrad}  we study in detail how the NP scalars change when projected onto a general tetrad. 
The fact that NP scalars are tetrad dependent is a well known result. In particular it is well-known that, in a  large class of alternative theories of gravity, the polarization content of a wave is an observer dependent quantity. See \cite{will_2018} for a pedagogical introduction. 

Once beyond geometric optics corrections are included, the total wave changes its propagation properties as diffraction effects are taken into account. 
Wave mechanics differs increasingly from geometric optics as the wavelength increases relative to the scale length of the medium inhomogeneities. The number of paths that can combine constructively increases and the rays that connect two points become blurred. In our description, this phenomenon manifests itself in the appearance of effective polarization modes  along the direction of the geometric optics ray. 
%
We emphasize that this does not mean that there are actual new physical degrees of freedom: everything is still expressed in terms of 4 real initial conditions, i.e.~we have only 2 propagating degrees of freedom. 


The results of this section show that beyond geometric optics lead to diffraction effects that smear the  polarization plane transverse to $k^\mu$,  and small new vector and scalar polarizations arise when projecting the wave onto a parallel transported tetrad. Different observers would measure different amount of extra-helicities modes, as explained in the appendix. This is a consequence of the fact that in the presence of diffraction the definition of a wave vector becomes an ambiguous concept. The issue of attempting to define one single propagation direction beyond geometric optics has been discussed recently in \cite{Harte:2018wni} (see Section 3.5) in the context of both electromagnetic and gravitational waves. It is shown that it is not necessarily meaningful to define a single propagation vector for finite-wavelength lensing as different phenomena may have different directions associated. In the next section we will explicitly compute the average direction of propagation of the wave's energy in the presence of diffraction. 



\section{Propagation of energy}\label{propofenergy}
 
 As we have previously discussed, when corrections beyond geometric optics are included, the null tetrad loses its precise physical meaning. In particular $k^\mu$ will not represent the direction of propagation of energy anymore.
Due to the absence of a geometrical definition of propagation, in this section, we study the effective propagation of energy of the wave as a physically meaningful quantity.
This quantity can be reconstructed by a direct inspection of the pseudo stress-energy momentum tensor of the wave.
 
 In the absence of curvature (sufficiently far from the lens), the energy momentum tensor of gravitational waves can be written as \cite{Maggiore:1900zz}
   \be
   t_{\mu\nu}=\frac{c^4}{32 \pi G} \partial_{\mu}h_{\alpha\beta}\partial_{\nu}h^{\alpha\beta}\,, 
   \ee
where we have reintroduced units of $c$ and $G$ to make contact with standard results in the literature. 
    Using (\ref{11}) and (\ref{12}) we can write the geometric optics contribution and the contribution beyond geometric optics as:
\be\label{expt}
   t_{\mu\nu}=\frac{c^4}{32 \pi G}\left[ \omega^2  t_{\mu\nu}^{(0)}+\omega  t_{\mu\nu}^{(1)}+\mathcal{O}(\omega^{0})\right]\,. 
   \ee
We calculate these terms and obtain:
\begin{align}
 t^{(0)}_{\mu\nu}=k_{\mu}k_{\nu}&\left[\sin^2\Phi \Re(\epsilon_{\alpha\beta}^{(0)}) \Re(\epsilon^{(0)\alpha\beta})\right.\nonumber\\
 &+\cos^2\Phi \Im(\epsilon_{\alpha\beta}^{(0)}) \Im(\epsilon^{(0)\alpha\beta})\nonumber\\
 &\left.-2\sin\Phi\cos\Phi \Re(\epsilon_{\alpha\beta}^{(0)}) \Im(\epsilon^{(0)\alpha\beta})\right]\,
\end{align}
and
\begin{align}
 t^{(1)}_{\mu\nu}=2k_{(\mu}&\left[-\sin\Phi\cos\Phi \Re(\epsilon_{\alpha\beta}^{(0)}) \partial_{\nu)}\Re(\epsilon^{(0)\alpha\beta})\right.\nonumber\\
 &+\cos^2\Phi \Im(\epsilon_{\alpha\beta}^{(0)}) \partial_{\nu)}\Re(\epsilon^{(0)\alpha\beta})\nonumber\\
  &-\sin^2\Phi \Re(\epsilon_{\alpha\beta}^{(0)}) \partial_{\nu)}\Im(\epsilon^{(0)\alpha\beta})\nonumber\\
    &\left.+\cos\Phi\sin\Phi \Im(\epsilon_{\alpha\beta}^{(0)}) \partial_{\nu)}\Im(\epsilon^{(0)\alpha\beta})\right]\nonumber\\
    +2k_{\mu}k_{\nu}&\left[\sin^2\Phi \Re(\epsilon_{\alpha\beta}^{(0)}) \Re(\epsilon^{(1)\alpha\beta})\right.\nonumber\\
 &+\cos^2\Phi \Im(\epsilon_{\alpha\beta}^{(0)}) \Im(\epsilon^{(1)\alpha\beta})\nonumber\\
  &-\sin\Phi\cos\Phi \Re(\epsilon_{\alpha\beta}^{(0)}) \Im(\epsilon^{(1)\alpha\beta})\nonumber\\
 &\left.-\sin\Phi\cos\Phi \Re(\epsilon_{\alpha\beta}^{(1)}) \Im(\epsilon^{(0)\alpha\beta})\right]\,.
\end{align}

Next, we define the effective energy momentum tensor, obtained averaging over several oscillations of the wave, as 
\be\label{average}
t^{\text{eff}}_{\mu\nu}=\frac{c^4}{32 \pi G}\langle \partial_{\mu}h_{\alpha\beta}\partial_{\nu}h^{\alpha\beta}\rangle\,, 
\ee
where $\langle\dots\rangle$  denotes a time average over several periods of the wave.
In our context, the fast oscillating part of the wave is driven by the eikonal phase $\Phi$ and  thus we average over $\Phi$ (equivalently, over several fast oscillations, at a fixed location). By doing this, we find in  the geometric optics limit the standard result (see e.g. \cite{Maggiore:1900zz}) 
\be
 t^{(0)\text{eff}}_{\mu\nu}=\frac{c^4}{64 \pi G}A^2 k_{\mu} k_{\nu}\,,
   \ee
   where $A$ is the amplitude of the polarization in geometric optics as defined in eq.\,(\ref{AAA}). 
   For the leading order corrections to geometric optics we get:
    \begin{align}\,\label{T1eff}
      t^{(1)\text{eff}}_{\mu\nu}&= \frac{c^4}{32 \pi G} k_{(\mu}\left[ \Im(\epsilon^{(0)\alpha\beta})\partial_{\nu)}\Re(\epsilon_{\alpha\beta}^{(0)})\right]\nonumber\\
    &-  \frac{c^4}{32 \pi G} k_{(\mu}\left[ \Re(\epsilon^{(0)\alpha\beta})\partial_{\nu)}\Im(\epsilon_{\alpha\beta}^{(0)})\right]\nonumber\\
      &+\frac{c^4}{32 \pi G}k_{\mu}k_{\nu}\Re\left(\epsilon_{\alpha\beta}^{(0)}\epsilon^{*\alpha\beta(1)}\right)\,.
   \end{align}
%
Since the polarization tensor in geometric optics has the form in (\ref{Epsilon0}), using the orthogonality property of the tetrad, eqs. (\ref{ortoI})-(\ref{ortoF}), we obtain that only the coefficients $\alpha_{\ell\ell}^{(1)}$ and $\alpha_{mm}^{(1)}$ are non-vanishing in eq. (\ref{T1eff}). We also observe that for a wave linearly polarized at emission,  $t^{(1)\text{eff}}_{\mu\nu}=0$ being $\Im(\epsilon_{\alpha\beta}^{(0)})=0=\Re(\epsilon_{\alpha\beta}^{(1)})$. In other words, for a linearly polarized wave corrections to the energy momentum tensor at one order beyond geometric optics cancel out once averaging over several periods of oscillation. This is due to the fact that, as mentioned in Sec.\,\ref{beyond}, for a linearly polarized wave the geometric optic solution is real and corrections at one order beyond geometric optics are dephased of $\pi/2$ with respect to the geometric optics order (i.e.~they are purely imaginary).

We can write the (effective) energy momentum tensor up to order $\omega$ in the following compact form 
 \be\label{interpretation}
 t^{\text{eff}}_{\mu\nu}=\frac{c^4}{64 \pi G}A^2 K_{\mu}  K_{\nu}\,,
   \ee
   where we have introduced the vector
   \be
K_{\mu}=k_{\mu}+\mathcal{K} k_{\mu}+\mathcal{V}_{\mu}\,,
    \ee
   with
   \begin{align}
   \mathcal{K} &=A^{-2}\Re\left(\epsilon_{\alpha\beta}^{(0)}\epsilon^{*\alpha\beta(1)}\right)\,,\\
   \mathcal{V}_{\mu}&=A^{-2}\left[ \Im(\epsilon^{(0)\alpha\beta})\partial_{\mu}\Re(\epsilon_{\alpha\beta}^{(0)})-\Re(\epsilon^{(0)\alpha\beta})\partial_{\mu}\Im(\epsilon_{\alpha\beta}^{(0)})\right]\,.
   \end{align}
For a point-like lens, an order of magnitude estimate of the size of these two corrections gives  $\mathcal{K}\sim \epsilon^{(1)}\sim (\lambda/b)(r_E/b)^2$ and $\mathcal{V}/k\sim \partial \epsilon ^{(0)}/k\sim \epsilon^{(1)}\sim (\lambda/b)(r_E/b)^2$, where we used eq.~(\ref{gaugeeq}) and (\ref{source2}).
    A tentative interpretation of this result is that, at first order beyond geometric optics, the real vector $K^{\mu}$, as opposite to $k^{\mu}$ gives the effective direction along which the average energy of the wave \emph{mainly} propagates.  We recall that here $\omega$ is just an expansion parameter, with no physical meaning. However, since it multiplies the phase $\Phi$ (see eq.\,(\ref{defomega})), flipping the sign of $\omega$ from $1$ to $-1$ corresponds to a flip of the helicity of the wave. It follows that, since the corrections  in eq.\,(\ref{interpretation}) are proportional to $\omega$, waves of opposite helicity do not feel the same effect when propagating on a curved background.  We expect this effect to be particularly important for waves propagating across a Kerr black hole. Indeed, in this case it is known that in the long wavelength limit gravitational waves of opposite helicity are scattered in a  different way, see e.g. \cite{Dolan:2008kf}. A similar result has been recently found in \cite{Dolan:2018ydp} for the case of electromagnetic waves.
An explicit computation and analysis of this effect for various astrophysical lenses will be addressed in the future.

\section{Discussion}\label{discussion}

\begin{table}[ht!]
	\begin{center}
		\begin{tabular}{ c| c| c }
			\hline
			Detector & Frequency (Hz) & Wavelength (pc) \\
			\hline
			LIGO/Virgo & $10^0-10^3$ & $10^{-8}-10^{-11}$ \\
			LISA & $10^{-4}-10^{-1}$ & $10^{-4}-10^{-7}$  \\
			IPTA & $10^{-9}-10^{-6}$ & $10^{-1}-10^{1}$  \\
			\hline
		\end{tabular}
		\caption{\label{bella}  Frequency and wavelength range of different GW observatories.}
	\end{center}
\end{table}

We have shown that beyond geometric optic corrections become important when the wavelength of the wave is of the order (or larger) than the  Schwarzschild radius of the lens. 
In table \ref{bella}, we mention relevant wavelengths for present and future GW observatories. The range $10^{-9}-10^{-6}$~Hz is covered by 

the International Pulsar Timing Array Consortium\footnote{{\tt http://www.ipta4gw.org}} (IPTA). Frequencies in the range $10^{-4}-10^{-1}$~Hz will be probed with the space-based LISA scheduled to be launched in 2034. Higher frequencies ($1-10^{3}$~Hz) are accessible with ground-based interferometers, including Advanced LIGO (aLIGO) \cite{2015CQGra..32g4001L} and Advanced Virgo (aVirgo) \cite{2015CQGra..32b4001A}, KAGRA interferometer which is expected to become operational by the end of 2019 and  LIGO India which is currently under construction. A third generation of ground-based interferometers, the Einstein Telescope\footnote{{\tt http://www.et-gw.eu}} (ET) and the Cosmic Explorer (CE) \cite{Evans:2016mbw} are in their design stages.

As a comparison, we also mention the values of Schwarzschild radii of different objects. Solar mass black holes have $r_S\sim 10^{-13}$pc, supermassive black holes have $r_S\sim10^{-9} $pc, and galaxies such as the Milky Way have $r_S\sim 10^{-2}$pc. In the LIGO frequency band wave effects are expected for waves passing in the vicinity of solar mass or some supermassive black holes. In the LISA band, wave effects are expected to appear when the lens is given by astrophysical objects in a wide mass range. 
For waves in the frequency range covered by IPTA, effects beyond geometric optics become relevant even when the lens is a galaxy (see e.g.~\cite{Cremonese:2018cyg} for an analysis on time delay including wave effects). 
When wave effects start becoming important, geometric optics may still remain a useful approximation and more accurate results can be obtained by including higher-order corrections, which provide insight into wave-optical phenomena that are not present in the eikonal limit.

In this article, we have analyzed  the propagation of gravitational waves on a curved background and we have proposed a perturbative method for studying corrections to the geometric optics limit. In particular, we have discussed the effects of beyond geometric optics corrections on the polarization tensor of the wave and we have illustrated the impact on the energy momentum tensor of the gravitational wave. We found that, in general, the wave beyond geometric optics gets diffractive effects that smear the polarization plane transverse to the geometric optics wave. As a result, the total wave exhibits longitudinal  components along  the geometric optics propagation vector, which is reflected on the appearance of additional (effective) vector and scalar polarization modes when projecting the Riemann tensor of the wave onto the standard eikonal 4-vector basis. In the absence of a natural definition of a single wave vector for the diffracted wave, we have analyzed the effective direction of propagation of the energy by direct inspection of the pseudo stress-energy momentum tensor (at first order beyond geometric optics). We found that the effective direction (obtained averaging over several \emph{fast} oscillations) of propagation of energy is misaligned with the eikonal direction of the wave.


We emphasize that in this paper we have made an initial study on how the propagation of a single plane wave gets affected by diffraction. Observationally, because of diffraction effects, the number of paths that can combine constructively increases and the rays that connect two points are the geometric optics rays plus diffracted components of other rays ignored in the geometric optics description. The fact that we have kept the tensor structure of the wave in the equations of motion, will allow us in the future to estimate what the net amplitude \emph{and} polarization of the detected wave will be for different lensing geometries.
The idea is to make use of a path integral approach similar to the one traditionally used to study diffraction of a scalar wave, see e.g.\,\cite{Takahashi:2003ix},  to study the diffraction pattern keeping track of the polarization structure of the wave. Ultimately, we will address the observability of the wave effects discussed in this article in different physical situations -- e.g.~for GW signals from binary systems detectable by LISA and lensed by a foreground stellar field -- and compute corrections to standard lensing quantities e.g. magnification and time-delay.\\




\textit{Acknowledgements ---} We are extremely grateful to P.~G.~Ferreira, W.~Hu, and L. Hui for their comments on this work. We also thank B.~Whiting, R.~Durrer, C.~Bonvin, C.~Pitrou, C.~Dalang, P.~Fleury, M.~Maggiore, R.~Wald, D.~E.~Holz, M.~Fishbach, R.~Essick, P.~Landry, and Z.~Doctor, and J.~M.~Ezquiaga for useful discussions. This project has received funding from the European Research Council (ERC) under the European Union's Horizon 2020 research and innovation program (grant agreement No 693024). ML was supported by the Kavli Institute for Cosmological Physics at the University of Chicago through an endowment from the Kavli Foundation and its founder Fred Kavli.

\appendix
\section{Tetrad Dependence}\label{App:Tetrad}
We have shown that due to diffraction effects, the Newman-Penrose scalars are not only characterized by tensor perturbations only $\Psi_4$, but instead all the other polarizations are excited. We  study here how this statement depends on the tetrad choice, i.e.~we analyze whether this statement is observer dependent. To this scope we consider a generalized Lorentz transformation of the tetrad and we check whether there exists a class of observers for whom the wave appears as a purely helicity-2 wave, i.e.~with $\Psi_4\neq 0$ and all other NP scalar vanishings. 

The most general transformation of the tetrad that preserves the orthonormal properties defined in eq.~(\ref{tetradeqns}) has 6 real free functions of time and space (generalization of Lorentz transformations in flat space), however two of them simply correspond to re-normalizations of the tetrad which are irrelevant for determining whether the NP scalars will vanish or not. See e.g.~\cite{will_2018} for a pedagogical derivation.  Explicitly, this transformation is:
\begin{equation}
k^{\mu'}= Ak^\mu, \quad n^{\mu'}=A^{-1}n^{\mu}, \quad m^{\mu'}=e^{i\Omega}m^\mu\,,
\end{equation}
with $A$ and $\Omega$ arbitrary real functions. In this case, the four relevant NP scalars will transform with a simple rescaling, as
\begin{equation}
\Psi_2'= \Psi_2,\; \Psi_3'= A^{-1}e^{-i\Omega}\Psi_3,\; \Psi_4'=A^{-2}e^{-i2\Omega}\Psi_4, \;\Phi_{22}'=A^{-2}\Phi_{22}.
\end{equation}
We therefore focus here on how the NP scalars transform under the remaining four free parameters of the general tetrad transformation, which is given by:
\begin{align}
&k^{\mu'}=k^\mu+|z_1|^2n^\mu+z_1^*m^\mu +z_1\ell^\mu\,,\\
& m^{\mu'}=m^\mu+z_1n^\mu+z_2k^\mu\,,\\
&n^{\mu'}= n^\mu +|z_2|^2k^\mu+z_2^*m^\mu +z_2\ell^\mu \,,
\end{align}
where $z_1$ and $z_2$ are two complex parameters.  
Under this transformation we find that all the NP scalars change as:
\begin{align}
&\Psi_2'=\Psi_2+\frac{2}{3}(z_1\Psi_3+z_1^*\Psi_3^*)+\frac{1}{6}(z_1^2\Psi_4+z_1^{*2}\Psi_4^*)+\frac{1}{3}|z_1|^2\Phi_{22}\,,\\
& \Psi_3'= \Psi_3+\frac{1}{2}(z_1^*\Phi_{22}+z_1\Psi_4)+ 3\Psi_2z_2^*\,,\\
&\Psi_4'=\Psi_4  +4\Psi_3z_2^*+6\Psi_2z_2^{*2},\\
&\Phi_{22}^{'}=\Phi_{22}+2(z_2\Psi_3+z_2^*\Psi_3^*) +6\Psi_2|z_2|^2\,.
\end{align}
The transformations driven by $z_1$ and $z_2$ are referred to as  class II and class I rotations, respectively. Class I rotations ($z_1$=0) correspond to the little group of Lorentz transformations that leaves the vector $k^{\mu}$ invariant, i.e.~they relate a class of observers that identify the same wave vector. Class I transformations are used to define a  quasi-Lorentz invariant classes of gravitational waves. Each class is labeled by the Petrov type of its non-vanishing Weyl tensor and the maximum number of non-vanishing amplitudes as seen by any observer \cite{will_2018}. We see that if the four NP scalars are non-vanishing, the complex transformation parameter $z_1$ is not sufficient to set all of them but $\psi_4$ to zero.
If we include the $z_2$ transformation, which transforms the wave vector, we see that there exists a special choice of $z_1$ and $z_2$ such that $\Psi_2'=\Psi_3'=\Phi_{22}'=0$: e.g.~the real part of $z_2$ is fixed to make $\Phi_{22}'$=0, the real part of $z_1$ to make $\Psi_2'=0$, and then the remaining two imaginary parts are fixed by the requirement that $\Psi_{3}'=0$. 
However, in this case, other projections of the Weyl that were originally vanishing and hence ignored in our analysis (e.g.~the so-called NP scalars $\Psi_0$ or $\Psi_1$), will now be excited when performing a class II transformation, see \cite{Carmeli:1975wq}. 

We therefore conclude that the identification of NP scalars in the presence of diffraction is an observer-dependent statement, as it is the case in alternative models of gravity  \cite{will_2018}. Nevertheless, we find that there is no choice of the tetrad such that only $\Psi_4\not=0$ when beyond geometric optics corrections are taken into account. 
The issue of attempting to define one single propagation direction beyond geometric optics has been discussed in \cite{Harte:2018wni} (see Section 3.5) in the context of both electromagnetic and gravitational waves, where it was also concluded that it is not necessarily meaningful to define a single propagation vector for finite-wavelength lensing as different phenomena may have different directions associated. For example, we defined one propagation vector in Section \ref{propofenergy} determining the average direction of propagation of  energy, but other observables may have other average directions associated to them.

\bibliographystyle{apsrev4-1}
\bibliography{RefModifiedGravity}

\end{document}